\documentclass[nofootinbib,prd,twocolumn,showpacs,showkeys,preprintnumbers]{revtex4}
\usepackage{hyperref,amssymb,amsmath,mathrsfs,bm,graphicx}
\usepackage{epsfig,graphics}
\usepackage{color}
\begin{document}
\title{The post--quasi-static approximation: An analytical approach to gravitational collapse}
\author{L. Herrera}
\email{lherrera@usal.es}
\affiliation{Instituto Universitario de F\'isica
Fundamental y Matem\'aticas, Universidad de Salamanca, Salamanca 37007, Spain}
\author{A. Di Prisco}
\email{alicia.diprisco@ucv.ve}
\affiliation{Escuela de F\'\i sica, Facultad de Ciencias, Universidad Central de Venezuela, Caracas 1050, Venezuela}
\author{J. Ospino}
\email{j.ospino@usal.es}
\affiliation{Departamento de Matem\'atica Aplicada and Instituto Universitario de F\'isica
Fundamental y Matem\'aticas, Universidad de Salamanca, Salamanca 37007, Spain}
\date{\today}
\begin{abstract}

A semi--numerical approach proposed many years ago for describing  gravitational collapse in the post--quasi--static approximation \cite{hjr,hjr2,hjr3,hjr4}, is modified  in order to avoid the numerical integration of the basic differential equations the approach is based upon. For doing that we have to impose some restrictions on the fluid distribution. More specifically, we shall assume   the vanishing complexity factor condition, which   allows for analytical integration of the pertinent differential equations and  leads to physically interesting models. Instead, we show that neither the homologous nor the quasi--homologous evolution are acceptable since they lead to geodesic fluids, which are unsuitable for being described in the post--quasi--static approximation. Also, we prove that, within this approximation,  adiabatic evolution also leads to geodesic fluids and therefore we shall  consider exclusively dissipative systems. Besides the vanishing complexity factor condition, additional information is required  for a full description of  models. We shall propose different strategies for obtaining such an information, which are based on observables quantities (e.g. luminosity and redshift), and/or heuristic mathematical ansatz. To illustrate the method, we present two models. One model is inspired in the well known Schwarzschild interior solution, and another one is inspired in Tolman VI solution.
\end{abstract}
\date{\today}
\pacs{04.40.-b, 04.20.-q, 04.25.-g}
\keywords{Relativistic fluids, gravitational collapse.}
\maketitle
\section{Introduction}
In the study of self--gravitating systems there are three possible regimes of evolution. The simplest one is the static (stationary when rotations are allowed) regime, which is characterized by the existence of a time--like Killing vector  forming  a vorticity--free congruence (in the stationary  case the congruence is not vorticity--free ).  In the coordinate system adapted to this congruence the metric and the physical variables are invariant with respect to translations along the  time axe. 

Next, we have the quasi--static regime (QSR), in which case   the system is assumed to evolve, but slowly enough, so that it can be considered to be in equilibrium at each moment (the TOV equation is satisfied at all times). This implies  that the fluid distribution  changes  on a time scale that is very long as compared to the  hydrostatic time scale \cite{astr1}--\cite{astr2} (sometimes this time scale is also referred to as dynamical time scale, e.g. \cite{astr3}). Thus , in this regime  the evolution of the fluid may be regarded as a sequence of static
models, where the time between any two states of equilibrium is neglected (see \cite{qs0,qs1,qs2} for applications).

The QSR applies to a large variety of scenarios due to the fact that 
the hydrostatic time scale is very small during many phases of the life of the star \cite{astr2}, e.g. it is of the order of $27$ minutes, $4.5$ seconds   and $10^{-4}$ seconds, respectively   for the Sun,  a white dwarf and   a neutron star of one solar mass and $10$ Km radius.

Finally, we have the dynamic regime where the system is out of equilibrium, meaning that the TOV equation is not satisfied. The system changes on a time scale which is smaller than the hydrostatic time scale.

All this having been said, the following question is in order: can we approach the non--equilibrium by means of successive approximations? Or, equivalently: Is there life between quasi--equilibrium and non--equilibrium?

As it has been proved in the past (see \cite{hjr,hjr2,hjr3,hjr4} and references therein) the answer to the above questions is affirmative (is some cases at least), the corresponding regime is called post--quasi--static (PQSR), and can be regarded as the closest, non--equilibrium, regime to QSR.  Before proceeding farther, some important remarks are in  order  

\begin{enumerate}
\item First of all it should be stressed that the main motivation to consider the PQSR is to have the possibility  to study, in the simplest possible way,  those aspects of the object directly related to the non--equilibrium situation, which for obvious reasons cannot be described within the QSR.
\item Since we are  assuming the fact that we can approach the non--equilibrium by means of successive approximations, it goes without saying that not  any self--gravitating fluid will satisfy this requirement. In particular is meaningless, from the physical point of view, to consider geodesic fluids in PQSR,  since these fluids are always in the full dynamic regime (the only interaction in  this case being  the gravitational one).
\item It also should be clear that unlike the two precedent regimes, there is not a unique definition for PQSR. Here we shall assume the definition proposed in \cite{hjr,hjr2,hjr3,hjr4}
\end{enumerate}

Let us now elaborate on the main motivation of  our endeavor with this work.

To provide an accurate description of the gravitational collapse of a supermassive star, including  the final fate of such process (naked singularities, black holes, anything else), the mechanism behind a type II supernova event  \cite{cw66}--\cite{s7} or the structure and evolution of the compact object  resulting from such a process \cite{e3}--\cite{e2}, is a task of utmost relevance.

We have available  three approaches  to study  the gravitational collapse in the context of general relativity. On the one hand,  numerical methods \cite{lehner}--\cite{font}, which   allow for including more realistic equations of state. Nevertheless,  the obtained results,
in general, may be  highly model dependent. Besides,  difficulties, associated to numerical solutions of partial differential equations in presence of shocks may complicate further the problem.

Alternatively,  one may resort to  analytical solutions to Einstein equations, which  are more suitable for  general discussions, and   may  be relatively simple to analyze,  still containing  some of the essential features of a realistic situation (see for example \cite{an1,an2,an3,an4,an5,an6,an7,an8,an9,an10,an11} and references therein). However, often they resort to heuristic assumptions  whose justification is unclear.

Between  the two aforementioned approaches, we have semi-numerical techniques which  may be regarded as a ``compromise'' between the analytical and numerical approaches. These techniques are based on the PQSR approximation mentioned above, and were developed in \cite{hjr,hjr2,hjr3,hjr4} (see also \cite{pqs1,pqs2}). 

This third approach,  allows to reduce the initial system of partial differential equations into a system of  ordinary differential equations (referred to as surface equations) for quantities evaluated at the boundary surface of the fluid distribution. 

The approach relies  on  a set of conveniently defined  variables (referred to as ``effective'' variables)  plus  an heuristic ansatz on the later, whose rationale and justification become intelligible within the context of the PQSR. 

So far, the above mentioned approach has been used by solving numerically  the surface equations. In this work we complement the  approach  with a sensible physical condition, allowing us to avoid numerical integration, resorting exclusively to analytical methods.  Such a condition appears to be the vanishing of the complexity factor, as defined in \cite{ps1,ps2}. Other plausible conditions such as the homologous \cite{ps2} and the quasi--homologous  \cite{epjc} conditions have been considered but were dismissed due to the facts that they, within the PQSR, lead to geodesic fluids.

Besides the vanishing complexity factor condition, we have to resort to additional sources of information  in order to obtain a full description of the collapsing system. The number of possible strategies for doing that is  very large. Here we emphasize, on the one hand, on conditions suggested by  observables such as the luminosity profile and the gravitational  redshift. On the other hand we propose some heuristic mathematical constrains, justified by  previous experience on finding  time--dependent  solutions to Einstein equations, or, simply,  by the fact that they allow a simple analytical integration.

The organization of the manuscript is as follows. In the next section we introduce the basic variables and definitions, as well as the Einstein and the transport equations.  In Section III, we detail the junction conditions with the exterior spacetime, which is Vaidya. The complexity factor and the homologous and quasi--homologous evolution are defined in Section IV. A review of the approach is outlined in Section V, and some examples are analyzed in Section VI.  Finally we include a discussion of the results and some concluding remarks in the last section.

\section{Basic variables and equations}
 \subsection{The metric}
 We consider a spherically symmetric distribution  of collapsing
fluid, bounded by a spherical surface $\Sigma$. The fluid is
assumed to be locally anisotropic (principal stresses unequal) and undergoing dissipation in the
form of heat flow (to model dissipation in the diffusion approximation). Physical arguments  to consider such fluid distributions in the study of gravitational collapse may be found in \cite{report}--\cite{DHN} and references therein.

Using comoving coordinates, we write the line element in the form
\begin{equation}
  ds^2=-A^2dt^2+B^2dr^2+R^2(d\theta^2+\sin^2\theta d\phi^2),
\label{metric}
\end{equation}
where $A$, $B$ and $R$ are functions of $t$ and $r$ and are assumed
positive. We number the coordinates $x^0=t$, $x^1=r$, $x^2=\theta$
and $x^3=\phi$.
\subsection{Energy--momentum tensor}
The matter energy--momentum tensor $T_{\alpha\beta}$ inside $\Sigma$
has the form
\begin{eqnarray}
T_{\alpha\beta}&=&(\mu +
P_{\perp})V_{\alpha}V_{\beta}+P_{\perp}g_{\alpha\beta}+(P_r-P_{\perp})K_{
\alpha}K_{\beta}\nonumber\\
&+&q_{\alpha}V_{\beta}+V_{\alpha}q_{\beta}, \label{3}
\end{eqnarray}
where $\mu$ is the energy density, $P_r$ the radial pressure,
$P_{\perp}$ the tangential pressure, $q^{\alpha}$ the heat flux, $V^{\alpha}$ the four-velocity of the fluid,
and $K^{\alpha}$ a unit four-vector along the radial direction.
 These quantities
satisfy
\begin{eqnarray}
V^{\alpha}V_{\alpha}&=&-1, \;\; V^{\alpha}q_{\alpha}=0, \;\; K^{\alpha}K_{\alpha}=1,\nonumber\\ \;\;
K^{\alpha}V_{\alpha}&=&0. \;\;\;\;\;\; \nonumber
\end{eqnarray}

Since we assume the metric (\ref{metric}) comoving then
\begin{eqnarray}
V^{\alpha}=A^{-1}\delta_0^{\alpha}, \;\;
q^{\alpha}=qB^{-1}\delta^{\alpha}_1, \;\;
K^{\alpha}=B^{-1}\delta^{\alpha}_1, \label{5bis}
\end{eqnarray}
where $q$ is a function of $t$ and $r$.

 \subsection{Kinematical variables}
 The four--acceleration $a_{\alpha}$ and the expansion $\Theta$ of the fluid are
given by
\begin{equation}
a_{\alpha}=V_{\alpha ;\beta}V^{\beta}, \;\;
\Theta={V^{\alpha}}_{;\alpha}, \label{4b}
\end{equation}
and its  shear $\sigma_{\alpha\beta}$ by
\begin{equation}
\sigma_{\alpha\beta}=V_{(\alpha
;\beta)}+a_{(\alpha}V_{\beta)}-\frac{1}{3}\Theta h_{\alpha \beta},\label{4a}
\end{equation}
where $h_{\alpha \beta}=g_{\alpha\beta}+V_{\alpha}V
_{\beta}
.$

We do not explicitly add bulk viscosity to the system because it
can be absorbed into the radial and tangential pressures, $P_r$ and
$P_{\perp}$, of the
collapsing fluid.

From  (\ref{4b}) with (\ref{5bis}) we have for the  four--acceleration and its scalar $a$,
\begin{equation}
a_1=\frac{A^{\prime}}{A}, \;\; a^2=a^{\alpha}a_{\alpha}=\left(\frac{A^{\prime}}{AB}\right)^2, \label{5c}
\end{equation}
where $a^\alpha= a K^\alpha$,
and for the expansion
\begin{equation}
\Theta=\frac{1}{A}\left(\frac{\dot{B}}{B}+2\frac{\dot{R}}{R}\right),
\label{5c1}
\end{equation}
where the  prime stands for $r$
differentiation and the dot stands for differentiation with respect to $t$.
With (\ref{5bis}) we obtain
for the shear (\ref{4a}) its non zero components
\begin{equation}
\sigma_{11}=\frac{2}{3}B^2\sigma, \;\;
\sigma_{22}=\frac{\sigma_{33}}{\sin^2\theta}=-\frac{1}{3}R^2\sigma,
 \label{5a}
\end{equation}
and its scalar
\begin{equation}
\sigma^{\alpha\beta}\sigma_{\alpha\beta}=\frac{2}{3}\sigma^2,
\label{5b}
\end{equation}
where
\begin{equation}
\sigma=\frac{1}{A}\left(\frac{\dot{B}}{B}-\frac{\dot{R}}{R}\right).\label{5b1}
\end{equation}
Then, the shear tensor can be written as
\begin{equation}
\sigma_{\alpha \beta}= \sigma \left(K_\alpha K_\beta - \frac{1}{3} h_{\alpha \beta}\right).
\label{sh}
\end{equation}

\subsection{Transport equations}
In the dissipative case we shall need a transport equation in order to find  the temperature distribution and its evolution. Assuming a causal dissipative theory (e.g. the Israel-- Stewart theory \cite{19nt,20nt,21nt}) the transport equation for the heat flux reads
\begin{eqnarray}
\tau h^{\alpha \beta}V^\gamma q_{\beta;\gamma}&+&q^\alpha=-k h^{\alpha \beta}\left(T_{,\beta}+Ta_\beta\right)\nonumber \\&-&\frac{1}{2}k T^2 \left(\frac{\tau V^\beta}{\kappa T^2}\right)_{;\beta} q^\alpha,
\label{tre}
\end{eqnarray}
where $k$,  $T$ and $\tau$ denote  thermal conductivity,  temperature and relaxation time respectively.

In the spherically symmetric case under consideration, the transport equation has only one independent component which may be obtained from (\ref{tre}) by contracting with the unit spacelike vector $K^\alpha$, it reads
\begin{equation}
\tau V^\alpha  q_{,\alpha}+q=-k \left(K^\alpha T_{,\alpha}+T a\right)-\frac{1}{2}k T^2\left(\frac{\tau V^\alpha}{\kappa T^2}\right)_{;\alpha} q.
\label{5}
\end{equation}

 \subsection{Field equations}
The Einstein field equations for the interior spacetime (\ref{metric}) can be written as
\begin{widetext}
 \begin{eqnarray}
 8\pi \mu  A^2=\left(2\frac{\dot B}{B} + \frac{\dot R}{R}\right)\frac{\dot R}{R}
 -\left(\frac{A}{B}\right)^2\left[ 2\frac{R^{''}}{R}+\left(\frac{R'}{R}\right)^2
 -2\frac{B'}{B}\frac{R'}{R}-\left(\frac{B}{R}\right)^2\right],\label{12}
 \end{eqnarray}

 \begin{equation}
 4\pi q AB=\left(\frac{\dot R'}{R}-\frac{\dot B}{B}\frac{R'}{R}-\frac{\dot R}{R}\frac{A'}{A}\right),
 \label{13}
 \end{equation}
 \begin{eqnarray}
 8\pi P_r B^2=-\left(\frac{B}{A}\right)^2\left[2\frac{\ddot R}{R}-\left(2\frac{\dot A}{A}-\frac{\dot R}{R}\right)
 \frac{\dot R}{R}\right]+\left(2\frac{A'}{A}+\frac{R'}{R}\right)\frac{R'}{R} - \left(\frac{B}{R}\right)^2,
 \label{14}
 \end{eqnarray}
 \begin{eqnarray}
 8\pi  P_\perp R^2=-\left(\frac{R}{A}\right)^2\left[\frac{\ddot B}{B}+\frac{\ddot R}{R} -\frac{\dot A}{A}
 \left(\frac{\dot B}{B}+\frac{\dot R}{R}\right)+\frac{\dot B}{B}\frac{\dot R}{R}\right]
 +\left(\frac{R}{B}\right)^2\left[\frac{A''}{A}+\frac{R''}{R}-\frac{A'}{A}\frac{B'}{B}+\left(\frac{A'}{A}-\frac{B'}{B}\right)
 \frac{R'}{R}\right].
 \label{15}
 \end{eqnarray}
\end{widetext}

At this point the following remark is in order: the knowledge of  $A(t,r)$, $B(t,r)$ and $R(t,r)$ casts the system above  in an algebraic system of four equations for the four unknown functions $\mu$,  $q$, $P_r$, and $P_{\perp}$ which, in such a case, can be obtained without further information.

\subsection{Mass and areal velocity}
Following Misner and Sharp \cite{ms64},
let us now introduce the mass function $m(t,r)$ (see also \cite{cm70}), defined by
\begin{equation}
m=\frac{R^3}{2}{R_{23}}^{23}
=\frac{R}{2}\left[\left(\frac{\dot R}{A}\right)^2-\left(\frac{R^{\prime}}{B}\right)^2+1\right].
 \label{17masa}
\end{equation}
It is useful to introduce  the proper time derivative $D_T$
given by
\begin{equation}
D_T=\frac{1}{A}\frac{\partial}{\partial t}, \label{16}
\end{equation}
and the proper radial derivative $D_R$,
\begin{equation}
D_R=\frac{1}{R^{\prime}}\frac{\partial}{\partial r}, \label{23a}
\end{equation}
where $R$ defines the areal radius of a spherical surface inside $\Sigma$ (as
measured from its area).

Using (\ref{16}) we can define the velocity $U$ of the collapsing
fluid  as the variation of the areal radius with respect to proper time, i.e.
\begin{equation}
U=D_TR. \label{19}
\end{equation}
Then (\ref{17masa}) can be rewritten as
\begin{equation}
E \equiv \frac{R^{\prime}}{B}=\left(1+U^2-\frac{2m}{R}\right)^{1/2}.
\label{20x}
\end{equation}

Using (\ref{12})-(\ref{14}) with (\ref{16}) and (\ref{23a}) we obtain from
(\ref{17masa})
\begin{eqnarray}
D_Tm=-4\pi\left(P_rU+q E\right)R^2,
\label{22Dt}
\end{eqnarray}
and
\begin{eqnarray}
D_Rm=4\pi\left(\mu+q \frac{U}{E}\right)R^2.
\label{27Dr}
\end{eqnarray}

Next, the three--acceleration $D_TU$ of an in-falling particle inside $\Sigma$ can
be obtained by using  (\ref{14}), (\ref{17masa})  and (\ref{20x}),
producing
\begin{equation}
D_TU=-\frac{m}{R^2}-4\pi P_r R
+E\frac{A^{\prime}}{AB}, \label{28pce}
\end{equation}
or
\begin{equation}
\frac{A^{\prime}}{A}=\frac{4\pi RB}{E}\left(\frac{D_TU}{4\pi R}+\frac{m}{4\pi R^3}+P_r\right). \label{29pce}
\end{equation}

Finally, from the Bianchi identities we obtain 
\begin{widetext}
\begin{eqnarray}
(\mu+ P_r)D_T U = -(\mu +  P_r)\left(\frac{m}{R^2}+ 4\pi P_rR\right)
-E^2\left[D_R P_r+ \frac{2}{R}( P_r- P_\perp)\right]-E\left[D_T q + 2 q \left(\frac{2U}{R}+\sigma\right)\right].\label{3m}
\end{eqnarray}
\end{widetext}
The  physical meaning of different terms in (\ref{3m}) has been discussed in detail in \cite{Hs}-\cite{DHN}. Suffice is to say in this point that  the first term on the right hand side describes the gravitational force term.

\section{The exterior spacetime and junction conditions}
Outside $\Sigma$ we assume we have the Vaidya
spacetime (i.e.\ we assume all outgoing radiation is massless),
described by
\begin{equation}
ds^2=-\left[1-\frac{2M(v)}{\rho}\right]dv^2-2d\rho dv+\rho^2(d\theta^2
+\sin^2\theta
d\phi^2) \label{1int},
\end{equation}
where $M(v)$  denotes the total mass,
and  $v$ is the retarded time.

The matching of the full non--adiabatic sphere  (including viscosity) to
the Vaidya spacetime, on the surface $r=r_{\Sigma}=$ constant, was discussed in
\cite{chan1}.

Now, from the continuity of the first  differential form it follows (see \cite{chan1} for details), 
\begin{equation}
A dt\stackrel{\Sigma}{=}dv \left(1-\frac{2M(v)}{\rho}\right)\stackrel{\Sigma}{=}d\tau, \label{junction1f}
\end{equation}
\begin{equation}
R\stackrel{\Sigma}{=}\rho(v), \label{junction1f2}
\end{equation}
and 
 \begin{equation}
\left(\frac{dv}{d\tau}\right)^{-2}\stackrel{\Sigma}{=}\left(1-\frac{2m}{\rho}+2\frac{d\rho}{dv}\right), \label{junction1f3}
\end{equation}
where $\tau$ denotes the proper time measured on $\Sigma$.

The continuity of the second differential form produces
\begin{equation}
m(t,r)\stackrel{\Sigma}{=}M(v), \label{junction1}
\end{equation}
and
\begin{widetext}
\begin{eqnarray}
2\left(\frac{{\dot R}^{\prime}}{R}-\frac{\dot B}{B}\frac{R^{\prime}}{R}-\frac{\dot R}{R}\frac{A^{\prime}}{A}\right)
\stackrel{\Sigma}{=}-\frac{B}{A}\left[2\frac{\ddot R}{R}
-\left(2\frac{\dot A}{A}
-\frac{\dot R}{R}\right)\frac{\dot R}{R}\right]+\frac{A}{B}\left[\left(2\frac{A^{\prime}}{A}
+\frac{R^{\prime}}{R}\right)\frac{R^{\prime}}{R}-\left(\frac{B}{R}\right)^2\right],
\label{j2}
\end{eqnarray}
\end{widetext}
where $\stackrel{\Sigma}{=}$ means that both sides of the equation
are evaluated on $\Sigma$ (observe a misprint in eq.(40) in \cite{chan1} and a slight difference in notation).

Comparing (\ref{j2}) with  (\ref{13}) and (\ref{14}) one obtains
\begin{equation}
q\stackrel{\Sigma}{=}P_r.\label{j3}
\end{equation}
Thus   the matching of
(\ref{metric})  and (\ref{1int}) on $\Sigma$ implies (\ref{junction1}) and  (\ref{j3}).

Also, we have
\begin{equation}
q\stackrel{\Sigma}{=}\frac{L}{4\pi \rho^2}, \label{20lum}
\end{equation}
where $L_\Sigma$ denotes   the total luminosity of the  sphere as measured on its surface and is given by
\begin{equation}
L \stackrel{\Sigma}{=}L_{\infty}\left(1-\frac{2m}{\rho}+2\frac{d\rho}{dv}\right)^{-1}, \label{14a}
\end{equation}
and where
\begin{equation}
L_{\infty} =-\frac{dM}{dv}\stackrel{\Sigma}{=} -\left[\frac{dm}{dt}\frac{dt}{d\tau}\left(\frac{dv}{d\tau}\right)^{-1}\right],\label{14b}
\end{equation}
is the total luminosity measured by an observer at rest at infinity.

The boundary redshift $z_\Sigma$ is given by
\begin{equation}
\frac{dv}{d\tau}\stackrel{\Sigma}{=}1+z,
\label{15b}
\end{equation}
with
\begin{equation}
\frac{dv}{d\tau}\stackrel{\Sigma}{=}\left(\frac{R^{\prime}}{B}+\frac{\dot R}{A}\right)^{-1}.
\label{16b}
\end{equation}
Therefore the time of formation of the black hole is given by
\begin{equation}
\left(\frac{R^{\prime}}{B}+\frac{\dot R}{A}\right)\stackrel{\Sigma}{=}E+U\stackrel{\Sigma}{=}0.
\label{17b}
\end{equation}
Also observe than from (\ref{junction1f3}), (\ref{14a})  and (\ref{16b}) it follows
\begin{equation}
L\stackrel{\Sigma}{=}\frac{L_\infty}{(E+U)^2},
\label{ju}
\end{equation}
and from (\ref{19}), (\ref{20x}), (\ref{junction1f3}) and (\ref{16b})
\begin{equation}
\frac{d\rho}{dv}\stackrel{\Sigma}{=}U(U+E).
\label{juf}
\end{equation}

\section{The complexity factor}
 The condition we shall impose on  our system in order to integrate analytically the ensuing differential equations, is the  vanishing of the complexity factor.  This is  a scalar function that has been proposed in order  to measure the degree of complexity of a given fluid distribution \cite{ps1, ps2}, and is related to the so called structure scalars \cite{sc}.

As shown in \cite{ps1, ps2} the complexity factor is identified with the scalar function $Y_{TF}$ which defines the trace--free part of the electric Riemann tensor (see \cite{sc} for details).

Thus,
let us define tensor $Y_{\alpha \beta}$ by
\begin{equation}
Y_{\alpha \beta}=R_{\alpha \gamma \beta \delta}V^\gamma V^\delta,
\label{electric}
\end{equation}
which may be expressed in terms of  two scalar functions $Y_T, Y_{TF}$, as
\begin{eqnarray}
Y_{\alpha\beta}=\frac{1}{3}Y_T h_{\alpha
\beta}+Y_{TF}\left(K_{\alpha} K_{\beta}-\frac{1}{3}h_{\alpha
\beta}\right).\label{electric'}
\end{eqnarray}

Then after lengthy but simple calculations, using field equations, we obtain (see \cite{ps2,epjc} for details)

\begin{equation}
Y_{TF}= -8\pi\Pi +\frac{4\pi}{R^3}\int^r_0{R^3\left(D_R {\mu}-3{q}\frac{U}{RE}\right)R^\prime d\tilde r}.
\label{Y}
\end{equation}
\\

In terms of the metric functions the scalar $Y_{TF}$ reads

\begin{widetext}
\begin{eqnarray}
Y_{TF}= \frac{1}{A^2}\left[\frac{\ddot R}{R} - \frac{\ddot B}{B} + \frac{\dot A}{A}\left(\frac{\dot B}{B} - \frac{\dot R}{R}\right)\right]+ \frac{1}{ B^2} \left[\frac{A^{\prime\prime}}{A} -\frac{A^{\prime}}{A}\left(\frac{B^{\prime}}{B}+\frac{R^{\prime}}{R}\right)\right] .
\label{itfm}
\end{eqnarray}
\end{widetext}

\subsection{The homologous and quasi--homologous evolution}
\label{sec:3}
Another set of possible conditions, which  might  be considered in order to avoid numerical integration, are conditions on the pattern of evolution.  

One of these conditions is represented by the homologous evolution ($H$). In \cite{ps2} it was assumed that the $H$ evolution describes the simplest mode of evolution of the fluid distribution. Such a condition is defined by 

\begin{equation}
 U=\tilde a(t) R ,\qquad \tilde a\equiv \frac{U_\Sigma}{R_\Sigma},
 \label{ven6}
 \end{equation}
 and
 \begin{equation}
\frac{R_I}{R_{II}}=\mbox{constant},
\label{vena}
\end{equation}
where $R_I$ and $R_{II}$ denote the areal radii of two concentric shells ($I,II$) described by $r=r_I={\rm constant}$, and $r=r_{II}={\rm constant}$, respectively. 
 
 These relationships  are reminiscent of the homologous evolution in Newtonian hydrodynamics \cite{astr1,astr2,astr3}.

The important point that we want to stress here is that,  in the relativistic regime,  (\ref{ven6}) does not imply (\ref{vena}).

Indeed, (\ref{ven6})  implies that for two comoving shells of fluids $I,II$ we have 
\begin{equation}
\frac{U_I}{U_{II}}=\frac{A_{II} \dot R_I}{A_I \dot R_{II}}=\frac{R_I}{R_{II}},
\label{ven3}
\end{equation}
which implies (\ref{vena}) only if the fluid is geodesic ($A={\rm constant}$). However,  in the non--relativistic regime, (\ref{vena}) always follows from the condition that  the radial velocity is proportional to the radial distance.

Another possible condition (less restrictive)  could be represented by the so called  ``quasi--homologous'' regime ($QH$), characterized by condition (\ref{ven6}) alone, which implies (see \cite{epjc} for details)

\begin{equation}
\frac{4\pi}{R^\prime}B  q+\frac{\sigma}{ R}=0.
\label{ch1}
\end{equation}

Thus the $H$ condition implies (\ref{vena}) and (\ref{ch1}), while the $QH$ condition only requires (\ref{ch1}). 

However  both conditions lead (within the PQSR) to geodesics fluids, which, as already mentioned, are physically without  interest.  

Indeed,   writing (\ref{13}) as 
\begin{equation}
4\pi qB=\frac{1}{3}(\Theta-\sigma)^{\prime}
-\sigma\frac{R^{\prime}}{R},\label{17a}
\end{equation}
and combining with condition  (\ref{ch1}), we obtain 
\begin{equation}
(\Theta-\sigma)^\prime=0,
\label{con5b}
\end{equation}
whereas, using (\ref{5c1}) and (\ref{5b1}) we get
\begin{equation}
\left(\Theta-\sigma\right)^\prime =\left(\frac{3}{A}\frac{\dot R}{R}\right)^\prime=0.
\label{con6b} 
\end{equation}

But  in the PQSR we have (see  equation (\ref{pce1}) in section 5.3 below) $R=\kappa(t) r$ where $\kappa$ is an arbitrary function of $t$,  producing  at once that 
\begin{equation}
A^\prime=0,
\label{con7}
\end{equation}
implying that  the fluid is geodesic, as it follows from (\ref{5c}). 

Thus from physical considerations we must exclude the $H$ or the $QH$ conditions for the mode of evolution.

We shall next, define mathematically the three regimes of evolution mentioned in the Introduction, in order to understand the rationale behind the proposed approach.
\section{Evolution regimes}
Let us now express the   three possible regimes of evolution, in terms of the metric and physical variables..
\subsection{Static regime}
In this case all time derivatives vanish, implying:
\begin{equation}
 q=U=\Theta=\sigma=0.
\label{1r}
\end{equation}

Since $B=B(r); A=A(r); R=R(r)$, reparametrizing $r$, we  may write the line element in the form:
\begin{equation}
ds^2=-A^2dt^2+B^2dr^2+r^2(d\theta^2+\sin^2\theta d\phi^2).
\label{2r}
\end{equation}

Thus, the ``dynamic equation (\ref{3m}) becomes the well known TOV equation of hydrostatic equilibrium for an anisotropic fluid
\begin{eqnarray}
P_r^{\prime}
+\frac{2}{r}(P_r-P_{\perp})=-\frac{(\mu+P_r)}{r(r-2m)}
(m
+4\pi P_r r^3).
\label{3r}
\end{eqnarray}
The Einstein equations in this case read:
\begin{eqnarray}
8\pi \mu A^2
=
-\left(\frac{A}{B}\right)^2\left[\left(\frac{1}{r}\right)^2
-2\frac{B^{\prime}}{Br}-\left(\frac{B}{r}\right)^2\right],
\label{4r} 
\end{eqnarray}
\begin{eqnarray}
8\pi P_r B^2 
=\left(2\frac{A^{\prime}}{A}+\frac{1}{r}\right)\frac{1}{r}-\left(\frac{B}{r}\right)^2,
\label{5r} 
\end{eqnarray}
\begin{eqnarray}
8\pi P_{\perp} r^2
=\left(\frac{r}{B}\right)^2\left[\frac{A^{\prime\prime}}{A}
-\frac{A^{\prime}}{A}\frac{B^{\prime}}{B}
+\left(\frac{A^{\prime}}{A}-\frac{B^{\prime}}{B}\right)\frac{1}{r}\right].\label{6r}
\end{eqnarray}

Also, for the mass function we have
\begin{equation}
m=\frac{r}{2}\left(1-\frac{1}{B^2}\right) \Rightarrow B^2=\left(1-\frac{2m}{r}\right)^{-1},
\label{7r}
\end{equation}

or
\begin{equation}
m=4\pi\int^{r}_{0} \mu r^2dr, \label{9r}
\end{equation}
and for the metric function $A$, we have from (\ref{29pce})
\begin{equation}
\ln\left(\frac{A}{A_\Sigma}\right)=\int_{r_\Sigma}^{r}\frac{(m+4\pi r^3 P_r)}{r(r-2m)} dr.\label{10r}
\end{equation}
The important point to keep in mind is that if  the radial dependence of $\mu$ and $P_r$ is known, the metric functions are determined  from  (\ref{7r}--\ref{10r}).

\subsection{Quasi-static regime (QSR)}
As mentioned before, in this regime the system is assumed to evolve, but sufficiently slow, so that it can be considered to be in equilibrium at each moment (Eq. (\ref{3r}) is satisfied). 

This implies for $U$,  the metric and the kinematical functions that

\begin{itemize}
\item The areal velocity $U$ and the kinematical variables are small, (of order $O(\epsilon)$, with $\vert\epsilon\vert<<1$) which  in turn implies that  dissipative variables and all first order time derivatives of  metric functions are also small, implying that  we shall neglect  terms of order $\epsilon^2$ and higher.

\item  From the above and the fact that the system always satisfies the equation of hydrostatic equilibrium, it follows from (\ref{3m}) that second time derivatives of  metric functions can be neglected.
\end{itemize}
Thus in QSR we have 
\begin{equation}
O(U^2) = {\dot A}^2 = {\dot B}^2 =
\dot A \dot B = \ddot R = \ddot B \approx0
\label{11r}
\end{equation}
and the radial dependence of the metric functions as well as that of physical variables is the same as in the static case. The only difference with the latter case being  that these variables depend upon time according to equation (\ref{13}).

\subsection{Post--quasi--static regime (PQSR)}
Let us now move one step forward into non--equilibrium and let us assume  that (\ref{3r}) is not satisfied. 

Then the question arises: What is the closest situation to QSR not satisfying eq. (\ref{3r})? Such a situation is described by what we call PQSR.

Since in both, the static and QSR regimes, the radial dependence of metric variables is the same, we shall keep that radial dependence as much as possible, but of course the time dependence  of  those variables is such that now (\ref{11r}) is not satisfied.

Then from the above we write
\begin{equation}
R=r\kappa(t), \label{pce1}
\end{equation}
where $\kappa$ is an arbitrary (dimensionless) function of $t$, to be determined later.

Taking into account (\ref{20x}) and (\ref{pce1}), we rewrite the metric as follows
\begin{equation}
ds^2=-A^2dt^2+ \kappa^2[E^{-2} dr^2+r^2(d\theta^2+{\sin}\theta^2 d\phi^2)].
\end{equation}

Next, defining the effective mass as
\begin{equation}
m_{eff}\equiv m-\frac{1}{2}R U^2,
\end{equation}
we obtain
\begin{equation}
E^2=1-\frac{2 m_{eff}}{R}.
\label{Eef}
\end{equation}

Then, equations (\ref{27Dr}) and (\ref{29pce}) can be written as
\begin{eqnarray}
\frac{1}{\kappa}m_{eff}^{\prime}&=&4\pi R^2 \mu_{eff},\label{drmeff}\\ \nonumber \\
\frac{1}{\kappa}(\ln{A})^{\prime} &=&\frac{4\pi R^2 P_{eff} + m_{eff}/R}{R-2m_{eff}},\label{drA}
\end{eqnarray}
with
\begin{eqnarray}
\mu_{eff}&=& \mu + \frac{ q U}{E} -\frac{U D_R U}{4\pi R} - \frac{U^2}{8\pi R^2},\label{dense}
\\ \nonumber \\
P_{eff}&=& P_r + \frac{D_T U}{4\pi R} + \frac{U^2}{8\pi R^2},\label{prese}
\end{eqnarray}
where we have followed  the terminology used in \cite{hjr2,hjr3,hjr4}  and  call  
$\mu_{eff}$ and $P_{eff}$ the ``effective density'' and the ``effective pressure'', respectively. The meaning of these variables will become clear in the discussion below, however we remark at this point that in the static and QSR cases, the effective variables coincide with the corresponding physical variables. (in what concerns their radial dependence).

Next, from (\ref{drmeff})--(\ref{prese}), with (\ref{pce1}) we may write
\begin{equation}
\frac{1}{\kappa^3} m_{eff}=\int^r_0 4\pi r^2 \mu_{eff}dr,\label{meff}
\end{equation}
\begin{equation}
\frac{1}{\kappa}\ln\left(\frac{A}{A_\Sigma}\right)=\int_{r_\Sigma}^r \left[\frac{4\pi R^3 P_{eff} + m_{eff}}{R(R-2m_{eff})}\right]
dr.\label{AE}
\end{equation}

From the above, we see  at once that if $R=\kappa(t) r$ and $\mu_{eff}$  have  the same radial dependence as $\mu$ in the static case, then the radial  dependence of $m_{eff}$ will be the same as in the static case. 

On the other hand, if besides the assumption above, we assume that $P_{eff}$ shares the same radial dependence as $P_{r}$ static, then it follows from (\ref{AE}) that $A$ shares the same radial dependence as in the static case.   

All these considerations provided the rationale for the  algorithm as exposed in \cite{hjr4}. Thus, the proposed method, starting from any interior (analytical) static spherically
symmetric (``seed'') solution to Einstein equations, leads to a system of ordinary differential equations for quantities evaluated at the boundary surface of the fluid distribution,
whose solution (numerical), allows for modeling, dynamic self-gravitating spheres, whose static limit is the original ``seed'' solution. 

In this work, motivated by our interest in resorting to purely analytical methods we shall modify the algorithm described in \cite{hjr4}. 

Specifically, the main steps of the formalism we propose may be summarized as follows.

\begin{enumerate}
\item Take an interior  (```seed'') solution to Einstein equations, representing a fluid distribution of matter in equilibrium, with a given 

$$\mu_{st}=\mu_{st}(r);\,\qquad\, P_{r st} = P_{r st}(r).$$

\item Assume that the $r$ dependence of the effective density   is the same as that of $\mu_{st}$,   and $R=r\kappa(t)$.  
\item Impose the vanishing complexity factor condition.

\item From the two conditions above we are able to determine the metric functions up to two arbitrary functions of $t$.

\item For these functions of $t$ one has the junction condition (\ref{j2}).

\item  In order to determine the remaining function and to integrate analytically  (\ref{j2})  we have a large number of possible strategies. Here we shall mention some of them,  which may be based  on the information obtained  from  the observables  of  the collapsing star. Such observables are the luminosity and the redshift. Alternatively we may assume additional heuristic constraints on some other physical variables, or ad hoc mathematical conditions based in previous works on gravitational collapse, or simply  justified by the fact that it allows a simple integration of (\ref{j2}). We list below some possible strategies of the kind mentioned above.
\begin{itemize}
\item Assuming a specific luminosity profile obtained from observations and using (\ref{14a}) or (\ref{14b}) we obtain  a relationship between  the two arbitrary functions of $t$ mentioned above, thereby reducing (\ref{j2}) to an ordinary differential equation for one variable.
\item Assuming a specific form for the evolution of the redshift we obtain again a relationship between  the two arbitrary functions of $t$
\item We may consider a specific pattern evolution of the areal radius of the star, or equivalently of its velocity ($U_\Sigma$). This could be useful if for example we want to check the possibility of a bouncing of the boundary surface.
\item Assuming different profiles of  either one of the two arbitrary functions of $t$, we can look for conditions   allowing the formation (or not) of a horizon, according to (\ref{17b}).
\end{itemize}

\end{enumerate}

\section{Modeling}
We shall now proceed to implement the approach for modeling  that we propose, and illustrate it by means of two examples.

Let us first write the general expressions for  the field equations and $Y_{TF}$. Using  (\ref{12})--(\ref{15}),  (\ref{itfm}) and (\ref{pce1}), we obtain 

\begin{equation}
8\pi\mu=\frac{1}{A^2}\left(\frac{2\dot B}{B}+\frac{\dot \kappa}{\kappa}\right)\frac{\dot \kappa}{\kappa}-\frac{1}{B^2}\left(\frac{1}{r}-\frac{2B^\prime}{B}\right)\frac{1}{r}+\frac{1}{r^2\kappa^2},
\label{muk}
\end{equation}
\begin{equation}
4\pi q=\frac{1}{AB}\left(\frac{\dot\kappa}{r\kappa}-\frac{\dot B}{rB}-\frac{A^\prime\dot\kappa}{A\kappa}\right),
\label{qk}
\end{equation}
\begin{eqnarray}
8\pi P_r=-\frac{1}{A^2}\left[\frac{2\ddot\kappa}{\kappa}-\left(\frac{2\dot A}{A}-\frac{\dot\kappa}{\kappa}\right)\frac{\dot\kappa}{\kappa}\right]\nonumber\\
+\frac{1}{B^2}\left(\frac{2A^\prime}{A}+\frac{1}{r}\right)\frac{1}{r}-\frac{1}{r^2\kappa^2},
\label{prk}
\end{eqnarray}
\begin{eqnarray}
8\pi P_\bot=-\frac{1}{A^2}\left[\frac{\ddot B}{B}+\frac{\ddot\kappa}{\kappa}-\frac{\dot A}{A}\left(\frac{\dot B}{B}+\frac{\dot\kappa}{\kappa}\right)+\frac{\dot B \dot\kappa}{B \kappa}\right]\nonumber\\
+\frac{1}{B^2}\left[\frac{A^{\prime\prime}}{A}-\frac{A^\prime B^\prime}{AB}+\left(\frac{A^\prime}{A}-\frac{B^\prime}{B}\right)\frac{1}{r}\right],
\label{ptk}
\end{eqnarray}
and 
\begin{eqnarray}
Y_{TF}=\frac{1}{A^2}\left[\frac{\ddot\kappa}{\kappa}-\frac{\ddot B}{B}+\frac{\dot A}{A}\left(\frac{\dot B}{B}-\frac{\dot\kappa}{\kappa}\right)\right]\nonumber\\
+\frac{1}{B^2}\left[\frac{A^{\prime\prime}}{A}-\frac{A^\prime}{A}\left(\frac{B^\prime}{B}+\frac{1}{r}\right)\right].
\label{Ttfk}
\end{eqnarray}

Let us first consider the $q=0$ case, which using (\ref{qk}) produces
\begin{equation}
\frac{1}{r}\left(\frac{\dot\kappa}{\kappa}-\frac{\dot B}{B}\right)-\frac{A^\prime\dot\kappa}{A\kappa}=0.
\label{qk0}
\end{equation}
Since at $r=0$, $A$ is different from zero,  we must impose

\begin{equation}
\frac{\dot\kappa}{\kappa}=\frac{\dot B}{B},\quad  \Rightarrow \rm B \; separable,
\label{so}
\end{equation}
 and 
\begin{equation}
\frac{A^\prime\dot\kappa}{A\kappa}=0, \quad \Rightarrow A=A(t), \; (\rm geodesic).
\label{at}
\end{equation}

Since the geodesic case in the PQSR should be dismissed by reasons exposed before,  we shall consider  exclusively dissipative systems.

Then since   $q\neq0$, it follows  from (\ref{qk})  that $B$ is separable
\begin{equation}
B(r,t)=\kappa(t) \beta(r),
\label{Bsep}
\end{equation}
here $\beta$ is an arbitrary dimensionless function of $r$, and 
\begin{equation}
4\pi q=-\frac{1}{A\kappa \beta}\left(\frac{A^\prime \dot\kappa}{A \kappa}\right).
\label{qd0}
\end{equation}

It is worth stressing that using (\ref{Bsep}) in  (\ref{5b1}) it follows at once that $\sigma=0$. Thus all our models will be shear--free.

Next, assuming $Y_{TF}=0$ we obtain from (\ref{Ttfk})
\begin{equation}
\frac{A^{\prime\prime}}{A^\prime}=\frac{\beta(r)^\prime}{\beta(r)}+\frac{1}{r},
\label{Y0qd0}
\end{equation}
whose solution reads
\begin{equation}
A=\alpha \int{\beta(r) rdr}+f(t),
\label{aqd0}
\end{equation}
where $f$ is arbitrary function of integration, and, by reparametrizying $t$,  another function of integration has been put equal to $\alpha=constant =1$, with dimensions $[1/r^2]$.

Then eqs.(\ref{muk})--(\ref{ptk}) take the form
\begin{equation}
8\pi\mu=\frac{1}{A^2}\frac{3 \dot \kappa^2}{\kappa^2}-\frac{1}{\beta^2 r \kappa^2}\left( \frac{1}{r}- \frac{2\beta^\prime}{\beta}\right)+\frac{1}{r^2\kappa^2},
\label{mukq}
\end{equation}
\begin{equation}
4\pi q=-\frac{\alpha r \dot \kappa}{A^2 \kappa^2},
\label{qkq}
\end{equation}
\begin{eqnarray}
8\pi P_r=-\frac{1}{A^2}\left(\frac{2\ddot\kappa}{\kappa}-\frac{2\dot f \dot \kappa}{A\kappa}+\frac{\dot \kappa^2}{\kappa^2}\right)\nonumber\\
+\frac{1}{\beta^2 r \kappa^2}\left(\frac{2 \alpha \beta r}{A}+\frac{1}{r}\right)-\frac{1}{r^2\kappa^2},
\label{prkq}
\end{eqnarray}
\begin{eqnarray}
8\pi P_\bot=-\frac{1}{A^2}\left(\frac{2 \ddot\kappa}{\kappa}-\frac{2\dot f \dot \kappa}{A \kappa}+\frac{\dot\kappa^2}{\kappa^2}\right)\nonumber\\
+\frac{1}{\beta^2 \kappa^2}\left(\frac{2 \alpha \beta}{A}-\frac{\beta^\prime}{r \beta}\right),
\label{ptkq}
\end{eqnarray}
where $A$ is given by (\ref{aqd0}).

Also, from (\ref{prkq}) and (\ref{ptkq})
\begin{equation}
8\pi(P_r-P_\bot)=\frac{1}{\beta^2 \kappa^2 r}\left(\frac{1}{r}+\frac{\beta^\prime}{\beta}\right)-\frac{1}{\kappa^2 r^2}.
\label{an}
\end{equation}

Using (\ref{pce1}) and (\ref{Bsep}) we can write
\begin{equation}
\mu_{eff}=\mu + \frac{q r\beta \dot\kappa}{A}-\frac{\dot\kappa^2}{8\pi A^2\kappa^2}\left(3-\frac{2\alpha r^2\beta}{A}\right),
\label{muef}
\end{equation}
\begin{equation}
P_{eff}=P_r+\frac{1}{4\pi A^2}\left(\frac{\ddot\kappa}{\kappa}-\frac{\dot\kappa \dot f}{\kappa A}\right)+\frac{\dot\kappa^2}{8\pi A^2 \kappa^2},
\label{pef}
\end{equation}
where $A$ is given by (\ref{aqd0}).

We shall now use the equations above to present some analytical models of collapsing objects. It should be stressed that the obtained models  are presented with the sole purpose of illustrating the method, and not to describe any specific astrophysical scenario.
\subsection{A model with homogenous effective energy--density}

\noindent The first model, is obtained by taking as our ``seed'' solution the well known Schwarzschild interior solution characterized by  homogeneous energy-density and isotropic pressure. 

Thus, assuming  $\mu_{eff}=F(t)$, where $F(t)$ is an arbitrary function with units $[1/r^2]$, we obtain from (\ref{meff}),
\begin{equation}
m_{eff}=\frac{4 \pi r^3 \kappa^3 F(t)}{3},
\label{mefF}
\end{equation}
and with (\ref{20x}) and (\ref{Eef}) we have
\begin{equation}
\frac{1}{r^2}\left(1-\frac{1}{\beta^2}\right)=\frac{8\pi \kappa^2 F(t)}{3},
\label{beF}
\end{equation}
then 
\begin{equation}
\beta^2=\frac{1}{1-cr^2},
\label{bet2}
\end{equation}
where c is a constant, with the same units as $F(t)$,  given by
\begin{equation}
c= \frac{8 \pi \kappa^2 F(t)}{3}.
\label{c}
\end{equation}
With this we have for $A$
\begin{equation}
A=f(t)-\frac{\alpha}{c}\sqrt{1-cr^2},
\label{aqde}
\end{equation}
and for the field equations
\begin{equation}
8\pi\mu=\frac{3 c^2 \dot\kappa^2}{\left(cf-\alpha\sqrt{1-c r^2}\right)^2\kappa^2}+\frac{3c}{\kappa^2},
\label{muc}
\end{equation}
\begin{equation}
4\pi q=-\frac{\alpha c^2 r \dot\kappa}{\left(cf-\alpha\sqrt{1-c r^2}\right)^2\kappa^2},
\label{qc}
\end{equation}
\begin{widetext}
\begin{eqnarray}
8\pi P_r=8\pi  P_\bot=-\frac{c^2}{\left(cf-\alpha \sqrt{1-c r^2}\right)^2} 
 \left[\frac{2\ddot\kappa}{\kappa}-\frac{2 c \dot f \dot\kappa}{\left(cf-\alpha \sqrt{1-c r^2}\right)\kappa}+\frac{\dot\kappa^2}{\kappa^2}\right]
+ \frac{2 c \alpha \sqrt{1-c r^2} }{\left(cf-\alpha\sqrt{1-c r^2}\right) \kappa^2}-\frac{c}{\kappa^2}.
\label{pc}
\end{eqnarray}
\end{widetext}
On the surface $\Sigma$, from (\ref{j2}) or (\ref{j3}) we obtain 
\begin{widetext}
\begin{eqnarray}
2\kappa \ddot\kappa-\frac{2c \dot f \kappa \dot\kappa}{\left(cf-\alpha \sqrt{1-c r^2}\right)}+\dot\kappa^2-2\alpha r\dot\kappa \stackrel{\Sigma}{=}4 \alpha f \sqrt{1-c r^2} - c f^2 - \frac{3\alpha^2 \left(1-c r^2\right)}{c}.
\label{sup}
\end{eqnarray}
\end{widetext}

Redefining $\alpha$ as
\begin{equation}
\alpha=\frac{c}{\sqrt{1-cr_\Sigma^2}},
\label{rea}
\end{equation}
equations (\ref{aqde})--(\ref{sup}) become
\begin{equation}
A=f-\sqrt{\frac{1-cr^2}{1-cr_\Sigma^2}},
\label{aqder}
\end{equation}

\begin{equation}
8\pi\mu=\frac{3 \dot\kappa^2}{\left(f-\sqrt{\frac{1-cr^2}{1-cr_\Sigma^2}}\right)^2\kappa^2}+\frac{3c}{\kappa^2},
\label{mucr}
\end{equation}
\begin{equation}
4\pi q=-\frac{ c r \dot\kappa}{\sqrt{1-cr_\Sigma^2}\left(f-\sqrt{\frac{1-cr^2}{1-cr_\Sigma^2}}\right)^2\kappa^2},
\label{qcr}
\end{equation}

\begin{widetext}
\begin{eqnarray}
8\pi P_r=8\pi  P_\bot=-\frac{1}{\left(f-\sqrt{\frac{1-cr^2}{1-cr_\Sigma^2}}\right)^2} 
 \left[\frac{2\ddot\kappa}{\kappa}-\frac{2  \dot f \dot\kappa}{\left(f-\sqrt{\frac{1-cr^2}{1-cr_\Sigma^2}}\right)\kappa}+\frac{\dot\kappa^2}{\kappa^2}\right]
+ \frac{2 c  \sqrt{\frac{1-c r^2}{1-c r_\Sigma^2}} }{\left(f-\sqrt{\frac{1-cr^2}{1-cr_\Sigma^2}}\right) \kappa^2}-\frac{c}{\kappa^2},
\label{pcr}
\end{eqnarray}
\end{widetext}
and
\begin{equation}
2\kappa \ddot\kappa-\frac{2 \dot f \kappa \dot\kappa}{\left(f-1\right)}+\dot\kappa^2-\frac{2 c r_\Sigma \dot\kappa}{\sqrt{1-cr_\Sigma^2}} = 4 f c - c f^2 - 3c.
\label{supr}
\end{equation}

Introducing the new variable
\begin{equation}
X\equiv\sqrt{c}(f-1),
\label{xr}
\end{equation}
(\ref{supr}) reads

\begin{equation}
2\kappa \ddot\kappa-\frac{2 \dot X \kappa \dot\kappa}{X}+\dot\kappa^2-\frac{2 c r_\Sigma \dot\kappa}{\sqrt{1-cr_\Sigma^2}} =-X^2+2\sqrt{c}X.
\label{supbr}
\end{equation}

Next, using (\ref{junction1f2}), (\ref{20lum}) and (\ref{qcr}) we obtain  for the luminosity on the surface

\begin{equation}
L_\Sigma=-\frac{cr_\Sigma^3 \dot \kappa}{\sqrt{1-cr_\Sigma^2}(f-1)^2},
\label{lus}
\end{equation}
or using (\ref{ju}), we obtain for the luminosity at infinity

\begin{equation}
L_\infty=-\frac{cr_\Sigma^3 \dot \kappa}{\sqrt{1-cr_\Sigma^2}(f-1)^2} \left(\sqrt{1-cr_\Sigma^2} + \frac{\dot \kappa r_\Sigma}{f-1}\right)^2.
\label{lusi}
\end{equation}

Also, observe that using (\ref{15b})  for this model, we obtain for the redshift at the boundary

\begin{equation}
z=\frac{(f-1)(\beta_\Sigma-1)-\dot \kappa r_\Sigma \beta_\Sigma}{f-1+\dot \kappa r_\Sigma \beta_\Sigma},
\label{red}
\end{equation}
and the time for the formation of a horizon is determined by the equation
\begin{equation}
\frac{\dot \kappa}{f-1}=-\frac{1}{\beta_\Sigma r_\Sigma}.
\label{bhg}
\end{equation}

Thus, the model is completely determined up to two functions of $t$ ($f$ and $\kappa$). As mentioned before, in order to determine these two functions we have a large number of possible strategies. Here we shall resort to heuristic   mathematical conditions, in order to fully determine the  system.

As a first example we shall assume  a heuristic mathematical condition on $\kappa$.
Thus, we shall next consider the  case  where  $\kappa$  has the linear form

\begin{equation}
\kappa=\kappa_0 t + \kappa_1,
\label{ls1}
\end{equation}
where $\kappa_0$ and $\kappa_1$ are arbitrary functions.
Then, introducing (\ref{ls1}) in (\ref{supbr}) we obtain
\begin{equation}
\frac{2 \dot f \kappa_0}{c (f-1)(f+b_1)(f+b_2)}=\frac{1}{\kappa_0 t + \kappa_1},
\label{ls2}
\end{equation}
whose solution is
\begin{eqnarray}
(f-1)^{b_1-b_2}(f+b_1)^{b_2+1}(f+b_2)^{-(b_1+1)}=\nonumber \\
 C (\kappa_0 t + \kappa_1)^{\frac{c(b_1+1)(b_2+1)(b_1-b_2)}{2\kappa_0^2}},
\label{ls3}
\end{eqnarray}
where $C$ is a constant and $b_1$ and $b_2$ have the following values
\begin{equation}
b_1=-2\pm\sqrt{1-\frac{\kappa_0\kappa_2}{c}},
\label{ls4}
\end{equation}
\begin{equation}
b_2=-2\mp\sqrt{1-\frac{\kappa_0\kappa_2}{c}},
\label{ls5}
\end{equation}
with
\begin{equation}
\kappa_2\equiv\kappa_0-\frac{2 c r_\Sigma}{\sqrt{1-cr_\Sigma^2}}.
\label{ls6}
\end{equation}

In order to obtain $f$ we have to solve the algebraic equation (\ref{ls3}), for any given set of constants. 

Thus, for example, for $b_1=0$, which implies $b_2=-4$, equation (\ref{ls3}) reads
\begin{eqnarray}
\frac{(f-1)^4}{f^3 (f-4)}=
 C (\kappa_0 t + \kappa_1)^{\frac{-6c}{\kappa_0^2}}.
\label{ls3bis}
\end{eqnarray}

In general for the particular solution (\ref{ls3}) the physical variables read

\begin{equation}
    8\pi \mu=\frac{1}{(\kappa_0 t+\kappa_1)^2}\left [\frac{3\kappa_0^2}{\left(f-\sqrt{\frac{1-cr^2}{1-cr^2_\Sigma}}\right)^2}+3c\right],
\end{equation}

\begin{equation}
    4\pi q=-\frac{c r \kappa_0}{\sqrt{1-cr^2_\Sigma}(\kappa_0 t+\kappa_1)^2\left(f-\sqrt{\frac{1-cr^2}{1-cr^2_\Sigma}}\right)^2},
\end{equation}
\begin{widetext}
\begin{eqnarray}
 8\pi P_r=8\pi P_\bot=\frac{2\kappa_0 \dot f}{(\kappa_0 t+\kappa_1)\left(f-\sqrt{\frac{1-cr^2}{1-cr^2_\Sigma}}\right)^3}- \frac{\kappa_0^2}{(\kappa_0 t+\kappa_1)^2\left(f-\sqrt{\frac{1-cr^2}{1-cr^2_\Sigma}}\right)^2}
\nonumber \\ + \frac{1}{(\kappa_0 t+\kappa_1)^2}\left [\frac{2c\sqrt{\frac{1-cr^2}{1-cr^2_\Sigma}}}{\left(f-\sqrt{\frac{1-cr^2}{1-cr^2_\Sigma}}\right)}-c\right],
\end{eqnarray}
\end{widetext}
\noindent  whereas for the luminosity we obtain
\begin{equation}
L_\Sigma=-\frac{c r_\Sigma ^3 \kappa_0}{\sqrt{1-cr^2_\Sigma}(f-1)^2}.
\end{equation}

Observe that in this particular case the   condition  for the formation of the horizon as implied by (\ref{bhg}) implies $f=constant$, which 
 obviously contradicts (\ref{ls3bis}). Thus no  black hole results from the evolution of such a model.

As a second example we shall next consider the particular case  $X=$constant, for which (\ref{supbr}) becomes
\begin{equation}
2\kappa \ddot\kappa+\dot\kappa^2-2\epsilon \dot\kappa =\xi,
\label{supbbr}
\end{equation}
where
\begin{equation}
\epsilon\equiv \frac{cr_\Sigma^2}{\sqrt{1-cr_\Sigma^2}}, \qquad \xi\equiv r_\Sigma^2(-X^2+2\sqrt{c}X),
\label{epr}
\end{equation}
and now dot denotes differentiation with respect to the dimensionless  variable $t/r_\Sigma$.

By introducing the variable
\begin{equation}
\dot \kappa=z \Rightarrow \ddot \kappa=\dot \kappa \frac{dz}{d\kappa}=z\frac{dz}{d\kappa},
\label{red1}
\end{equation}
the equation above becomes
\begin{equation}
2\kappa\frac{dz}{d\kappa}+\frac{1}{\kappa}(z^2-2\epsilon z)=\frac{\xi}{\kappa},
\label{red2}
\end{equation}
whose solution reads
\begin{equation}
z\equiv \dot \kappa=\frac{\xi^{1/2} \sqrt{\kappa+h}}{\sqrt{\kappa}},
\label{red3}
\end{equation}
where

\begin{equation}
h=\frac{2}{\gamma}\left[\ln{\kappa} \pm \sqrt{1+\gamma \kappa^2}\mp \ln{\left \vert\frac{1+\sqrt{1+\gamma \kappa^2}}{\kappa \gamma^{1/2}}\right\vert }\right],
\label{red4}
\end{equation}
and $\gamma$ is an arbitrary constant.

We shall not elaborate further on these  models, since the resulting expressions are too cumbersome, and  our sole purpose  here is to illustrate the way of using the proposed formalism, and not describe any specific  astrophysical scenario.
\subsection{A model obtained from Tolman VI as seed solution}
\noindent Our next model is inspired in the well known    Tolman $VI$ solution \cite{Tolm}, whose equation of state for large values of $\mu$ approaches that for a highly  compressed Fermi gas. 

Thus we assume
\begin{equation}
\mu_{eff}=\frac{g(t)}{r^2},\label{mueffVI}
\end{equation}
where $g$ is an arbitrary (dimensionless) function of $t$.
\noindent Using the above expression in  (\ref{drmeff}) it follows

\begin{equation}
    m_{eff}=4\pi\kappa ^3 g(t)r, \label{meffVI}
\end{equation}
\noindent and replacing  (\ref{meffVI}) into (\ref{Eef}) we obtain

\begin{equation}
    \frac{1}{\beta^2}=1-8\pi\kappa^2 g(t)=1-c, \label{betaVI}
\end{equation}
where $c$ and $\beta$ are dimensionless constants.

Then using (\ref{pce1}), (\ref{Bsep}), (\ref{aqd0}), (\ref{betaVI}), and redefining the constant $\alpha$ as 

\begin{equation}
  \alpha=\frac{2\sqrt{1-c}}{r_\Sigma^2} \label{betaVII},
\end{equation}
the metric variables for this model read
\begin{eqnarray}
    A&=&f(t)+\left(\frac{r}{r_\Sigma}\right)^2,\\
    B&=&\frac{\kappa}{\sqrt{1-c}}=\beta \kappa,\\
    R&=&\kappa (t) r,
\end{eqnarray}
and the expressions for the physical variables are

\begin{eqnarray}
    8\pi\mu&=&\frac{3\dot \kappa ^2}{\kappa ^2(\frac{r^2}{r^2_\Sigma}+f)^2} +\frac{\beta^2-1}{r^2\kappa ^2\beta^2},\\
    4\pi q&=&-\frac{2r \dot \kappa}{\kappa^2 \beta r^2_\Sigma (\frac{r^2}{r^2_\Sigma}+f)^2},\\
    8\pi P_r&=&-\frac{1}{(\frac{r^2}{r^2_\Sigma}+f)^2}
    \left [ \frac{2 \ddot \kappa}{\kappa}-\frac{2 \dot f \dot \kappa}{\kappa(\frac{r^2}{r^2_\Sigma}+f)}+ \frac{\dot \kappa^2}{\kappa^2} \right]\nonumber \\&+&\frac{4}{\kappa^2 \beta^2 r^2_\Sigma (\frac{r^2}{r^2_\Sigma}+f)}-\frac{\beta^2 -1}{\ \beta^2 \kappa^2 r^2},\\
    8\pi (P_r-P_\bot)&=&-\frac{\beta^2-1}{\beta^2\kappa^2 r^2},
\end{eqnarray} 
\noindent  whereas the junction condition, the luminosity and the redshift read 
\begin{equation}\label{jt}
 2 \ddot \kappa \kappa-\frac{2\dot f \dot \kappa  \kappa}{(f+1)}+\dot\kappa^2-4\frac{ \dot \kappa}{\beta r_\Sigma}= \frac{4}{\beta^2 r^2_\Sigma}\left(f+1\right) -\frac{\beta^2-1}{\beta^2 r^2_\Sigma}\left(f+1 \right)^2
\end{equation}

\begin{equation}
L_\Sigma=-\frac{2  r_\Sigma \dot \kappa}{\beta(f+1)^2},
\label{lust}
\end{equation}

\begin{equation}
L_\infty=-\frac{2 r_\Sigma \dot \kappa (f+1+\beta r_\Sigma \dot \kappa)^2}{\beta^3 (f+1)^4},
\label{lusit}
\end{equation}
and 
\begin{equation}
z=\frac{(f+1)(\beta-1)-\dot \kappa r_\Sigma \beta}{f+1+\dot \kappa r_\Sigma \beta},
\label{red}
\end{equation}
implying that  the time for the formation of a horizon is determined by the equation
\begin{equation}
\frac{\dot \kappa}{f+1}=-\frac{1}{\beta r_\Sigma}.
\label{bhgt}
\end{equation}

It would be convenient to write (\ref{jt}) in terms of the dimensionless variable $\bar t\equiv t/r_\Sigma$, it reads

\begin{equation}
 2 \ddot \kappa \kappa-\frac{2\dot f \dot \kappa  \kappa}{(f+1)}+\dot\kappa^2-4\frac{ \dot \kappa}{\beta}= \frac{4}{\beta^2 }\left(f+1\right) -\frac{(\beta^2-1)}{\beta^2}\left(f+1 \right)^2 \label{jtb},
\end{equation}
where now dots denote derivatives with respect to $\bar t$.

As in the precedent case we have a large number of possible strategies to obtain the two functions of $t$ determining the whole system. Thus we could consider for example the $f=constant$ case, or the assumption of the linearity of $\kappa$. In both cases the procedure is very similar as in the preceding case. Instead, we shall propose a different approach here. 

Specifically we shall split (\ref{jtb})  in two equations, as follows 

\begin{equation}
 2 \ddot \kappa \kappa+\dot\kappa^2-4\frac{ \dot \kappa}{\beta}= 0 \label{jtb1},
\end{equation}

\begin{equation}
-\frac{2\dot f \dot \kappa  \kappa}{(f+1)}= \frac{4}{\beta^2 }\left(f+1\right) -\frac{(\beta^2-1)}{\beta^2}\left(f+1 \right)^2 \label{jtb2}.
\end{equation}

Equation (\ref{jtb1}) may be integrated producing
\begin{equation}
\frac{ -2 \omega b \sqrt{\kappa}+b^2 \kappa +2 \omega^2 \ln{(\omega+b \sqrt{\kappa})}}{b^3}=t+\gamma,
\label{jtb3}
\end{equation}
where $\omega$ and $\gamma$ are two integration constants and $b\equiv4/\beta$.

Solving the above transcendental equation for $\kappa$ and feeding the result back into (\ref{jtb2}) we obtain $f$.

Once the functions of time are determined, we have to resort to a transport equation (e.g. (\ref{tre})) in  order to find  the distribution and evolution of the temperature. 

As in the previous example, the resulting expressions are too burdensome  and not very illuminating, so we shall  not elaborate further on them.

\section{Discussion and Conclusions}

We have proposed an analytical approach to describe spherical collapse within the context of PQSR. To avoid the numerical integration of differential equations appearing in the algorithm put forward in \cite{hjr,hjr2,hjr3,hjr4}, we have assumed the vanishing complexity factor as the cornerstone of the proposed method.  As far as we are aware, this is the first approach for modeling gravitational collapse  which includes, both, the PQSR and the vanishing complexity conditions. Doing so, starting with a given ``seed'' static analytical solution  to the Einstein equation,  we are led to a situation where the whole system is determined by two arbitrary functions of $t$. These functions are related through the junction condition (\ref{j2}). For the additional information required to obtain the above mentioned functions, we have presented a list of possible strategies, based on either information obtained from observables such as luminosity and gravitational redshift,   or from ad hoc heuristic mathematical conditions imposed on the system. It goes without saying that the presented list is not exhaustive, and much more possibilities can be considered. In this work, and with the sole purpose to illustrate the method,  we have resorted to heuristic mathematical restrictions. It must be clear that  the full  potential of the approach may only be deployed when the missing information is provided by either of the observables mentioned above.  Although this last issue remains one the most important pending question regarding our approach, it is out of the scope of this manuscript.

Invoking the vanishing complexity factor as the main assumption behind the proposed approach is not arbitrary, and its rationale becomes intelligible when we remind that the complexity factor has been shown to be a good measure of the degree of complexity of a fluid distribution. Thus, assuming such a condition we ensure that  we are dealing with the ``simplest'' fluid distributions available within the PQSR,  in concord with one of the main goals of our endeavor consisting  in describing gravitational collapse in its 
simplest possible way.

There is an additional argument reinforcing the assumption of vanishing complexity factor within the context of PQRS. Indeed, as we have seen,  all models obtained with the approach here presented, are necessarily shear--free. On the other hand, as shown in  \cite{stsf}, the shear--free condition is unstable in the presence of pressure anisotropy and/or dissipation.
However, writing the complexity factor in  terms of kinematical variables as 
\begin{equation}
Y_{TF}= \frac{a^\prime}{B} -a \frac{R'}{RB}+ a^2 -\frac{\dot{\sigma}}{A} - \frac{\sigma^2}{3} - \frac{2}{3} \Theta \sigma ,
\label{shevp}
\end{equation}
it can be shown that the vanishing of the complexity factor implies the stability of the shear--free condition in the geodesic case (seen \cite{stsf} for details). In the non--geodesic, static, case the combination of the first three terms on the right of (\ref{shevp}) must be equal to zero if we assume the vanishing of the complexity factor, implying in its turn that such combination must remain non--vanishing but small (bounded) in the PQSR. In such a case we may safely conclude that  the quasi--stability of $\sigma=0$ is ensured (see the discussion between Eqs.(63) and (67) in \cite{stsf}).

Conditions  on the complexity of the pattern of evolution such as $H$ and $QH$,  appear to be too strong and have to be excluded since they  lead to geodesic
fluids, which as mentioned before are physically incompatible with the very idea behind the PQSR.

Also, the adiabatic condition implies that the fluid is geodesic, accordingly we have considered exclusively dissipative fluids.

In order to illustrate the method we have presented two models. One is based on the interior Schwarzschild solution  as the ``seed'' solution, whereas the other is inspired in the well known Tolman VI solution. The purpose of these calculations was to show how the algorithm works. In order  to provide the missing information we have resorted to some mathematical ansatz. We would like to emphasize once again  that the optimal path to display the power  of the presented method  would be  to supply such information through  physical data obtained from astrophysical observations, among which the luminosity and the gravitational redshift appear to be the most relevant. We harbor the hope that some of our colleagues will be able to succeed in such endeavor.

\end{document}